\documentclass[journal]{IEEEtran}

\usepackage{amssymb,amsthm,amsmath}
\usepackage{graphicx}                       
\usepackage{epstopdf}
\usepackage[usenames,dvipsnames]{xcolor}    
\usepackage{siunitx}                        
\usepackage{subfigure}                      
\usepackage{fancyhdr}                       
\usepackage{balance}                        
\usepackage[hidelinks]{hyperref}            

\graphicspath{{../}{./figures/}}
\UseRawInputEncoding

\newtheorem{definition}{Definition}
\newtheorem{proposition}{Proposition}
\usepackage{booktabs}
\usepackage{threeparttable}
\usepackage{multirow}
\usepackage{url}
\usepackage{algorithm}
\usepackage[noend]{algpseudocode}

\usepackage{enumitem}
\usepackage{tabularray}
\usepackage{booktabs}
\usepackage{stfloats}
\usepackage{cite}

\usepackage{xcolor,soul,framed} 
\colorlet{shadecolor}{yellow}

\hypersetup{
colorlinks=true,
linkcolor=blue,
citecolor=blue
}

\begin{document}
\title{Physics-Augmented Data-EnablEd Predictive Control for Eco-driving of Mixed Traffic Considering Diverse Human Behaviors} %
%
%
\author{Dongjun Li,
        Kaixiang Zhang,
	Haoxuan Dong,
        Qun Wang,
        Zhaojian Li,
        and Ziyou Song 
\thanks{Manuscript received 24 April 2024; accepted 30 Jan 2023. Recommended by xxx. (Corresponding author: Ziyou Song.)}
\thanks{D. Li, H. Dong, and Z. Song are with the Department of Mechanical Engineering, National University of Singapore, 117575 Singapore, Singapore (e-mail: dongjun.li@u.nus.edu; donghaox@foxmail.com; ziyou@nus.edu.sg).}
\thanks{K. Zhang and Z. Li are with the Department of Mechanical Engineering, Michigan State University, East Lansing, MI 48824, USA. (e-mail: zhangk64@msu.edu; lizhaoj1@msu.edu).}
\thanks{Q. Wang is with the Department of Mechanical Engineering, National University of Singapore, 117575 Singapore, Singapore, and also with the School of Mechanical Engineering, Nanjing University of Science and Technology, Nanjing 210094, China (e-mail:wangqun0516@gmail.com).}
\thanks{Color versions of one or more figures in this article are available at https://xxxx}
\thanks{Digital Object Identifier xxx}}
%
%

\markboth{IEEE Transactions on Control Systems Technology}%
{Shell \MakeLowercase{\textit{et al.}}: Bare Demo of IEEEtran.cls for IEEE Journals}



\maketitle 
\begin{abstract}
Data-driven cooperative control of connected and automated vehicles (CAVs) has gained extensive research interest as it can utilize collected data to generate control actions without relying on parametric system models that are generally challenging to obtain. Existing methods mainly focused on improving traffic safety and stability, while less emphasis has been placed on energy efficiency in the presence of uncertainties and diversities of human-driven vehicles (HDVs). In this paper, we employ a data-enabled predictive control (DeePC) scheme to address the eco-driving of mixed traffic flows with diverse behaviors of human drivers. Specifically, by incorporating first-order physics, we propose the Physics-Augmented DeePC approach to handle substantial system variations and uncertainties. Then, a novel optimization framework, avoiding conventional reference trajectories during the receding optimization process, is proposed to further reduce the holistic energy consumption of mixed traffic flows. Simulation results demonstrate the effectiveness of our approach in accurately capturing random human driver behaviors and reducing holistic energy consumption, while ensuring driving safety and traffic efficiency. Furthermore, the proposed optimization framework achieves substantial reductions in energy consumption, i.e., average reductions of $4.14\%$, $4.97\%$ and $9.29\%$ when compared to the benchmark algorithms.
\end{abstract}

%
\IEEEpeerreviewmaketitle

\section{Introduction}
%
%
%
%
\IEEEPARstart{T}{he} emergence of connected and automated vehicles (CAVs) has paved the way towards next-gen vehicular systems with enhanced energy efficiency, reduced emissions, and improved mobility \cite{contreras2017internet,abboud2016interworking,zhu2021multi,dong2023flexible}. Eco-driving, as a key advance in promised CAV technology, aims to improve the safety and mobility of traffic systems while simultaneously minimizing the overall energy consumption \cite{wegener2021automated}. The eco-driving can be achieved by incorporating information from surrounding traffic and road conditions into real-time control and optimization schemes to effectively adjust the speed and acceleration of controllable vehicles. As CAVs and human-driven vehicles (HDVs) are anticipated to coexist on roads in the near future, it is necessary to design eco-driving control of CAVs in a mixed traffic flow. This is challenging as HDVs often exhibit intricate, nonlinear behaviors that are difficult to characterize and cooperate with \cite{wang2021leading,li2023human}. In addition, it has been demonstrated both theoretically and experimentally that incorporating diverse driving behaviors of HDVs into CAV controller design can enhance mixed traffic performance~\cite{zheng2020smoothing,stern2018dissipation,orosz2016connected}. Therefore, the development of CAV controllers needs to carefully address the uncertainties caused by HDVs such that the whole traffic system can operate in a safe and efficient manner.

The mixed traffic flow is a highly complex human-in-the-loop system. Although various approaches have been proposed to characterize the behavior of human drivers, such as the intelligent driver model (IDM) \cite{treiber2000congested}, the optimal velocity model (OVM) \cite{bando1995dynamical}, and their variants \cite{herman1959traffic,gazis1961nonlinear}, accurately identifying the diverse and non-linear driving behaviors exhibited by human drivers remains a significant challenge, which cannot be easily tackled by model-based methods \cite{hu2022distributed}. As such, researchers have investigated novel approaches that can bypass the identification of human driver behaviors, by utilizing model-free or data-driven methods such as reinforcement learning (RL) \cite{wang2023adaptive, zhu2021multi} and adaptive dynamic programming \cite{huang2020learning} to develop CAV control strategies that can optimize vehicle operations and enhance mobility. By considering the complex interactions between CAVs and HDVs, these strategies have demonstrated promising energy-saving performance for both CAVs and HDVs. For instance, Zhu \textit{et al}. \cite{zhu2021multi} proposed a model predictive control (MPC) method for CAVs in mixed traffic flows, which incorporates an integrated data-driven car-following model to predict behaviors of HDVs. Simulation results demonstrated the effectiveness in improving energy efficiency while maintaining driving safety and control robustness. Wang \textit{et al.} \cite{wang2023adaptive} employed the RL method to optimize the control actions of a CAV and learned the behavior of the following HDV, thereby reducing the total energy consumption in a holistic framework. 
However, these methods generally have a heavy computational burden and may struggle to handle constraints critical to vehicle safety. 

Data-EnablEd Predictive Control (DeePC) recently emerged as a promising model-free optimal control paradigm that directly uses input-output data to achieve safe and optimal control of unknown systems \cite{coulson2019data}. Instead of relying on a parametric system model, DeePC is a non-parametric approach that leverages Willems' fundamental lemma \cite{willems2005note} to directly predict future system trajectories \cite{coulson2019regularized}. This approach is effective for handling real-time uncertainty and disturbances \cite{dorfler2022bridging}. Furthermore, DeePC allows for the incorporation of input/output constraints to meet safety requirements \cite{zhang2022data}. Notably, DeePC is equivalent to sequential system identification and MPC for deterministic linear time-invariant (LTI) systems and has demonstrated promising control performance and lower computation cost for nonlinear and non-deterministic systems \cite{coulson2022data, zhang2022data, wang2023deep}. For instance, Wang \textit{et al.} \cite{wang2023deep} proposed a Data-EnablEd Predictive Leading Cruise Control algorithm for CAVs to smooth mixed traffic, which has been expanded to address large-scale problems \cite{wang2023distributed}. However, their research did not consider the diverse driving behaviors of human drivers or the optimization of energy consumption, which are vital aspects of CAV operation. In real-world traffic scenarios, human drivers may exhibit vastly different driving styles \cite{wang2023adaptive}, ranging from aggressive to mild, potentially leading to the failure of DeePC due to the lack of a pre-known driving environment with random surrounding HDVs. Furthermore, current eco-driving strategies rely on the tracking of pre-defined reference trajectories (set points tracking), yet their optimality is hard to ensure in uncertain environments. Essentially, adherence to a reference trajectory limits the optimization search space, potentially missing more efficient routes that arise in dynamic real-time conditions. \cite{hu2023generic,yamashita2016reference}. Hence, the consideration of diverse driving behaviors and the incorporation of energy consumption optimization in DeePC are still missing but worth careful investigation. 

In this paper, we develop an eco-driving control strategy to achieve safe and energy-efficient control of the mixed traffic flow, considering diverse human driver behaviors. The main contributions are highlighted as follows:
\begin{itemize}
    \item First, we proposes a Physics-Augmented Data-EnablEd Predictive Control (PA-DeePC) strategy to handle substantial system variations or uncertainties. By incorporating partial system physics, the prediction accuracy of system dynamics is significantly improved. 
    \item Second, we study the eco-driving problem of mixed traffic flows considering diverse human driver behaviors. Particularly, a novel eco-driving framework is proposed to address the restricted energy optimization problems of adopting pre-defined speed and spacing references when facing diverse HDVs' behaviors. Our method could further reduce the holistic energy consumption, while respecting all safety-related constraints.
    \item Last but not least, comprehensive simulation results are provided to demonstrate the effectiveness and robustness of the developed approaches under extensive real-world traffic conditions involving random HDVs.
\end{itemize}


\section{Preliminaries}\label{Preliminaries}

This section outlines a non-parametric representation to capture the behavior of linear systems and the formulation of DeePC. The subscript $_k$ is employed to indicate discrete representation, while $t$ signifies continuous representation.

\subsection{Non-Parametric Representation of System Behavior}
Consider a discrete LTI system, the state-space equation is the most commonly used description in system identification, where $A \in \mathbb{R}^{n \times n}$, $B \in \mathbb{R}^{n \times m}$, $C \in \mathbb{R}^{p \times n}$, $D \in \mathbb{R}^{p \times m}$, $x_{k} \in \mathbb{R}^{n}$, $u_{k} \in \mathbb{R}^{m}$ and $y_{k} \in \mathbb{R}^{p}$ represent the state, input vector and output vector, respectively.
\begin{equation}
	\label{eq_ssfunction}
    \left\{
	\begin{aligned}
		& x_{k+1} = Ax_{k} + Bu_{k}, \\
		& y_{k} = Cx_{k} + Du_{k}.
	\end{aligned}
    \right.
\end{equation}

The dynamic behavior of this LTI system can be represented by a sufficiently rich collection of its input/output data revealed by Willems' fundamental lemma \cite{willems2005note} . Specifically, this lemma begins by collecting a length $T\in \mathbb{N}$ sequence of input/output trajectory data $u^{d}_{[1,T]} = \mathrm{col}(u^{d}_{1}, \cdots, u^{d}_{T})\in \mathbb{R}^{mT}$, and $y^{d}_{[1,T]} = \mathrm{col}(y^{d}_{1}, \cdots, y^{d}_{T})\in \mathbb{R}^{pT}$ with a sampling interval of $\Delta t \in \mathbb{R}$. Then the Hankel matrix of depth $L$ for the inputs $u^{d}_{[1,T]}$ can be defined as
\begin{equation} \label{eq_Hankel_U}
\begin{aligned}
    \mathcal{H}_{L} (u^d_{[1,T]}) &=
	\begin{bmatrix}
		{u}^{d}_{1}            & {u}^{d}_{2}              & \cdots  & {u}^{d}_{T-L+1} \\
		{u}^{d}_{2}            & {u}^{d}_{3}              & \cdots  & {u}^{d}_{T-L+2} \\
		\vdots                 & \vdots                   & \ddots  & \vdots      \\
		{u}^{d}_{L}            & {u}^{d}_{L+1}            & \cdots  & {u}^{d}_{T}
	\end{bmatrix}.
 \end{aligned}
\end{equation}
Similarly, the Hankel matrix for the outputs can be defined as $\mathcal{H}_{L} (y^d_{[1,T]})$. Let $T_{\mathrm{ini}}$, $N\in \mathbb{Z}$, and $L=T_{\mathrm{ini}}+N$. The Hankel matrices $\mathcal{H}_{L}(u^d_{[1,T]})$ and $\mathcal{H}_{L}(y^d_{[1,T]})$ are divided into two parts, as follows:
\begin{equation}
    \begin{bmatrix}
        U_{p} \\ U_{f}
    \end{bmatrix} = \mathcal{H}_{L}(u^d_{[1,T]}), \qquad
    \begin{bmatrix}
        Y_{p} \\ Y_{f}
    \end{bmatrix} = \mathcal{H}_{L}(y^d_{[1,T]}),
\end{equation}
where $U_p$ and $U_f$ contain the past and future input data, respectively. 
Similarly, $Y_p$ and $Y_f$ represent the past and future output data, respectively. 
\\
\textbf{System behavior representation.} Motivated by Willems' fundamental lemma \cite{willems2005note} and the DeePC formulation \cite{coulson2019data}, we have the following result: Let $k > 0$ be the current time step. Then, we define the past control input sequence of length $T_{ini}$ as $u_{ini}=col(u_{k-T_{ini}},u_{k-T_{ini}+1},\cdots,u_{k-1}) \in \mathbb{R}^{m}$, and the predicted control input sequence within a time horizon of length $N$, denoted as $u=col(u_{k},u_{k+1},\cdots,u_{k+N-1}) \in \mathbb{R}^{n}$. Similarly, the past and predicted output sequences $y_{ini}$ and $y$ can also be defined.

\begin{definition}
(Persistently Exciting, in brief PE condition). A signal sequence $ u^{d} _{[1,T]}$ is persistently exciting of order $L (L \leq T) $ if the Hankel matrix $\mathcal{H} _L(u^{d}_{[1,T]})$ is of full row rank.
\end{definition}

\begin{proposition} 
Willems' fundamental lemma states that if the pre-collected input sequence $u^d_{[1,T]}$ is persistently exciting of order $ T_{ini}+N+n$ with $n$ being the dimension of the system states ($n$ can be chosen as an upper bound of state dimension), then at each time step, the patched trajectory $col(u_{ini},y_{ini},u,y)$ of the past $T_{ini}$ steps is generated from the system when it is spanned by $(U_p,Y_p,U_f,Y_f)$, that is, there exists a vector $g \in \mathbb{R}^{T-T_{ini}-N+1}$ such that \cite{wang2023deep}:
\begin{equation}
	\label{eq_g}
	\begin{bmatrix}
		U_p \\
		Y_p \\
		U_f \\
		Y_f \\
	\end{bmatrix}
    g = 
    \begin{bmatrix}
    	u_{ini} \\
    	y_{ini} \\
    	u       \\
    	y
    \end{bmatrix}.
\end{equation}
If $T_{ini} \geq$ lag of system, $y$ is uniquely determined from (\ref{eq_g}), $\forall (u_{ini},y_{ini},u)$. Therefore, the control input $u$ and the system output $y$ over the $N$-step prediction horizon can be obtained.
\end{proposition}

\begin{definition}
(Minimal Length of $T$). In order for $u^{d}_{[1,T]}$ to be PE of order $L$, it must have $T \geq (m+1)L+n-1 $, that is, the input signal sequence $u^{d}_{[1,T]}$ should be sufficiently rich and long as to excite the system yielding an output sequence that is representative for the system behavior \cite{coulson2019data}.
\end{definition}

\subsection{Data-EnablEd Predictive Control}

However, Eq. (\ref{eq_g}) is only valid for deterministic LTI systems. For nonlinear or non-deterministic systems \cite{coulson2019data}, we introduce a slack variable $\sigma _y \in \mathbb{R}^{(n+m)T_{ini}}$ for the system past output to ensure the feasibility of the equality constraint \cite{coulson2019data,coulson2019regularized,coulson2021distributionally}, yielding the following optimization problem

\begin{subequations}
	\label{eq_Regularized}
	\begin{align}
		\underset{g,u,y,\sigma_y}{\text{min}}  & \sum_{k=0}^{N-1}(\Vert y_{k} \Vert^2_Q + \Vert u_k \Vert^2_R) + \lambda_g\Vert g \Vert^2_2 + \lambda_y\Vert \sigma_y \Vert^2_2, \label{eq_regularizedcost}  \\
		\text{s.t.}  \quad &
		\begin{bmatrix}
			U_p \\
			Y_p \\
			U_f \\
			Y_f 
		\end{bmatrix}
		g = 
		\begin{bmatrix}
			u_{ini} \\
			y_{ini} \\
			u       \\
			y          
		\end{bmatrix}
		+
		\begin{bmatrix}
			0        \\
			\sigma_y \\
			0        \\
			0
		\end{bmatrix} ,                  \label{eq_regularizedlemma}        \\
		& u_k \in \mathcal{U}_{k}, \forall k \in \{0,\cdots,N-1\},  \label{eq_regularizedconst1}\\
		& y_k \in \mathcal{Y}_{k}, \forall k \in \{0,\cdots,N-1\},  \label{eq_regularizedconst2} 
	\end{align}
\end{subequations}
where $\mathcal{U}_{k}$ and $\mathcal{Y}_{k}$ represent the input and output constraints at time step $k$, respectively.  In Eq. (\ref{eq_Regularized}), the slack variable $\sigma_y$ is penalized with a weighted two-norm penalty function, and the weight coefficient $\lambda_y >0$ can be chosen sufficiently large such that $\sigma_y \ne 0$ only if the equality constraint is infeasible. In addition, a two-norm penalty on $g$ with a weight coefficient $\lambda_g >0$ is also incorporated to avoid over-fitting when noise-corrupted data samples are present. 
\subsection{Physics-Augmented DeePC}
Considering a LTI system with some first order physics or prior knowledge (e.g., velocity and acceleration), the physics can be generally represented as
\begin{equation}
	\label{eq_1st_order_physics}
    \left\{
	\begin{aligned}
		& \acute{x}_{k+1} = \acute{a}\acute{x}_{k} + \acute{b}\acute{u}_{k}, \\
		& \acute{y}_{k} = \acute{c}\acute{x}_{k} + \acute{d}\acute{u}_{k},
	\end{aligned}
    \right.
\end{equation}
where all variables ($\acute{x}_{k},\acute{y}_{k},\acute{x}_{k},\acute{a},\acute{b},\acute{c},\acute{d} $) are scalars, and generally $\acute{d}=0$. Hence, we can obtain the following relationship:
\begin{equation}
	\label{eq_physics_Hankel}
	\acute{y}_{k+1} =  \acute{c}(\underset{\acute{x}_{k+1}}{\underbrace{\acute{a}\acute{x}_{k} + \acute{b}\acute{u}_{k}}}) + \underset{0}{\underbrace{\acute{d}\acute{u}_{k+1}}} = \acute{a}\acute{y}_{k} + \acute{c}\acute{b}\acute{u}_{k}.
\end{equation}
Eq. (\ref{eq_physics_Hankel}) indicates partial physics between inputs and outputs. By incorporating this physics, we proposed the Physics-Augmented DeePC. When accounting for the presence of inaccurate physics, we define the physics residual as
\begin{equation}
	\label{eq_adaptive_physics}
	\epsilon_{k} = \acute{y}_{k+1} - \acute{a}\acute{y}_{k} - \acute{c}\acute{b}\acute{u}_{k}.
\end{equation}
Eq. (\ref{eq_adaptive_physics}) can be incorporated into DeePC (\ref{eq_Regularized}) through two methods. Eq. (\ref{eq_adaptive_physics}) can be treated as an equality constraint $(\epsilon_{k} = 0)$ when accurate physics are present, and as an inequality constraint $(\epsilon_{k} \in \mathcal{E}(k))$ when inaccurate physics are present. Alternatively, it can be added as a penalized cost term $\Vert \epsilon (k) \Vert ^{2}_{M}$, where the parameter's magnitude is contingent on the physical accuracy. 
By incorporating such suitable and accurate physics into DeePC, the non-parametric model can produce more accurate results, as the exploration among data sequences can be simplified by the physics.

\section{System modeling}\label{Foundation}
In this section, we begin by providing a concise overview of the linearized model used to describe the system dynamics of mixed traffic flows. Subsequently, we introduce a precise and computationally efficient energy consumption model that serves as the basis for optimizing energy consumption. Lastly, we discuss the techniques employed for modeling human driver diversity. 

\begin{figure*}[!ht]
        \centering
	\includegraphics[width=6.8in]{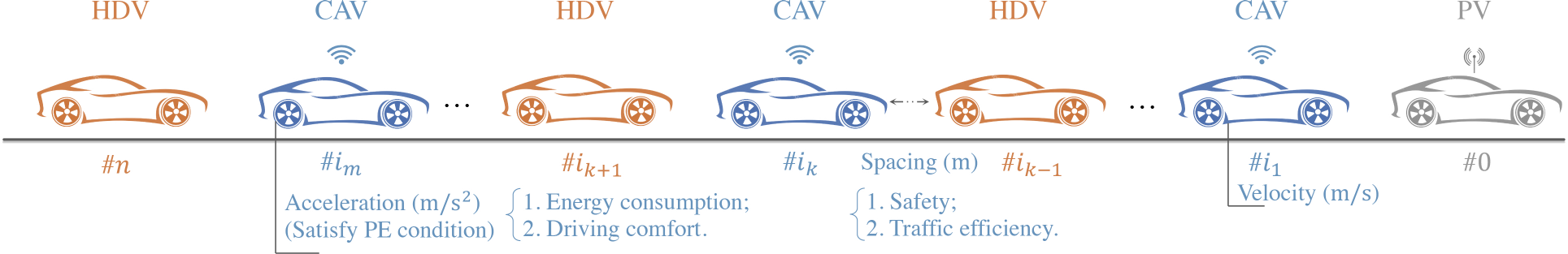}
	\caption{Schematic of the mixed traffic flows. The preceding vehicle is indexed as $0$. The subsequent $n$ vehicles comprise $m$ CAVs and $n-m$ HDVs with indeterminate driving dynamics.}
	\label{fig_Scenario}
\end{figure*}

The focus of this study is a single-lane mixed traffic flow of $n+1$ individual vehicles, as illustrated in Fig. \ref{fig_Scenario}. The preceding vehicle (PV) is indexed as 0. The system under consideration comprises $m$ CAVs and $n-m$ HDVs, with the set of CAV indices denoted as $S = \{1,2,\ldots,m\}$, the set of all vehicle indices as $\Omega = \{1,2,\ldots,n\}$, and the set of HDV indices as $\Omega \setminus S$. The index $i$ is used to represent every vehicle in the mixed traffic flows. In this study, we assume that HDV velocities and spacing are observable using V2X communications technology. Note that measuring each HDV's equilibrium spacing in mixed traffic flow is non-trivial due to diverse driving behaviors, which can result in HDV's spacing being unknown.

\subsection{Linearized Mixed Traffic Model}

The nonlinear system of the mixed traffic can be linearized according to the study by Jin \textit{et al}.\cite{jin2016optimal}. A class of continuous-time car-following models (e.g., OVM, and IDM) can be written in the following form to describe the car-following behavior of vehicle $i$.
\begin{equation}
	\label{eq_Generalform}
    \left\{
	\begin{aligned}
		\dot{s}_{i}(t) & = v_{i-1}(t) - v_{i}(t), \\
		\dot{v}_{i}(t) & = F(s_{i}(t),\dot{s}_{i}(t),v_{i}(t)),
	\end{aligned}
    \right.
\end{equation}
where $s_{i}(t)$ represents the relative spacing between vehicle $i$ and its preceding vehicle $i-1$; $\dot{s}_{i}(t)=v_{i-1}(t)-v_{i}(t)$ denotes the relative velocity between vehicle $i$ and its preceding vehicle $i-1$; and $v_{i}(t)$ stands for the velocity of vehicle $i$. Remark that $F(\cdot)$ can represent either OVM or IDM.

The linearization of the mixed traffic system is carried out around the equilibrium point. Assuming that each HDVs try to maintain the uniform traffic flow equilibrium velocity $v_{i}(t)\equiv v^{*}$ and the corresponding equilibrium spacing $s_{i}(t)\equiv s^{*}$, we can define the spacing and velocity perturbations, i.e., error states for velocity and spacing,

\begin{equation}
	\label{eq_errorstate}
	\widetilde{s}_{i}(t) = s_{i}(t) - s^{*},\qquad  \widetilde{v}_{i}(t) = v_{i}(t) - v^{*},
\end{equation}
where the estimated equilibrium velocity $v^{*}$ and the corresponding equilibrium spacing $s^{*}$ can be calculated as \cite{wang2023deep}
\begin{equation}
	\label{eq_equilibrium}
    \begin{aligned}
         v^{*} = \frac{1}{T_{ini}}\sum_{i=k-T_{ini}}^{k-1} v_{0}(i),\quad
         s^{*} = \frac{s_{gap}+ T_{gap}v^{*}}{\sqrt{1-(\frac{v^{*}}{v_{d}})^{4}}},
    \end{aligned}
\end{equation} \\
where $s_{gap}$ is the minimum gap, $T_{gap}$ is the (bumper to bumper) time headway to the preceding vehicle, and $v_{d}$ is the desired velocity. Then Eq. (\ref{eq_Generalform}) is linearized and the longitudinal dynamics model for each HDV is obtained and can be represented by 
\begin{equation}
	\label{eq_HDVdynamics}
	\left\{
		\begin{aligned}
			& \dot{\widetilde{s}}_{i}(t) = \widetilde{v}_{i-1}(t) - \widetilde{v}_{i}(t) ,      \\ 
			& \dot{\widetilde{v}}_{i}(t) = \alpha _{1}\widetilde{s}_{i}(t) - \alpha _{2}\widetilde{v}_{i}(t) + \alpha _{3}\widetilde{v}_{i-1}(t),
		\end{aligned}
	\right.
	  \quad i \in \Omega \setminus S,
\end{equation}
where $\alpha _{1} = \frac{\partial F}{\partial s}, \alpha _{2} = \frac{\partial F}{\partial \dot{s}} - \frac{\partial F}{\partial v}, \text{and } \alpha _{3} = \frac{\partial F}{\partial \dot{s}}$. More linearization details can refer to \cite{jin2016optimal}.
Similarly, the acceleration is used as the control input, then the longitudinal dynamics of each CAV can be denoted as
\begin{equation}
	\label{eq_CAVdynamics}
	\left\{
		\begin{aligned}
			& \dot{\widetilde{s}}_{i}(t) = \widetilde{v}_{i-1}(t) - \widetilde{v}_{i}(t),       \\ 
			& \dot{\widetilde{v}}_{i}(t) = u_{i}(t),
		\end{aligned}
	\right.
	\qquad i \in S.
\end{equation}
Finally, the state-space equation for mixed traffic flows can be obtained based on Eqs. (\ref{eq_HDVdynamics}) and (\ref{eq_CAVdynamics}). Please refer to \cite{wang2021leading} for more details. Then, by collecting the accelerations of all CAVs as system inputs $u^{d}_{[1,T]}$, and the velocity and spacing perturbations of all following vehicles as system outputs $y^{d}_{[1,T]}$, we construct the input Hankel matrix $\mathcal{H}_{L}(u^d_{[1,T]})$ and the output Hankel matrix $\mathcal{H}_{L}(y^d_{[1,T]})$, respectively.

\subsection{Energy Consumption Model}
To reduce computational complexity, an approximate and differentiable energy consumption model in the form of a polynomial expression is developed using experimental data in \cite{wang2023adaptive} to denote vehicle traction power
, as provided by
\begin{equation}
	\label{eq_polynomial}
	P_{tot}(t) = \sum_{i=0}^{3} \sum_{j=0}^{2}p_{ij}v^{i}(t)a^{j}(t),
\end{equation}
where $p_{ij}$ are constant parameters \cite{wang2023adaptive}, $v(t)$ is the vehicle velocity, and $a(t)$ is the acceleration. Here, we mix up the notation a little bit and simply use $i$, $j$ to represent power values. 

Considering the maximum order in Eq. (\ref{eq_polynomial}) exceeds 2, instead of a convex function. For the convenience of optimization problem solving, an order-reduction simplification is made through the incorporation of past speed into the higher-order term (e.g., $v(t)^{3}\approx \bar{v}v(t)^{2}$) to obtain an estimated power. In this way, the estimated total power can be expressed as a convex function, as shown below.
\begin{equation}
	\label{eq_estimatedPower}
	\begin{aligned}
		\widetilde{P}_{tot}(t) & = (p_{30}\bar{v}+p_{20})v(t)^{2} + (p_{12}\bar{v}+ p_{02})a(t)^{2} + (p_{21}\bar{v} \\
		& + p_{11})v(t)a(t)  + p_{10}v(t) + p_{01}a(t) + p_{00}.
	\end{aligned}
\end{equation}

\begin{figure}[!htpb]
     \centering
     \subfigure[Absolute error $P_{tot}-\widetilde{P}_{tol}$]{
         \includegraphics[width=0.225\textwidth]{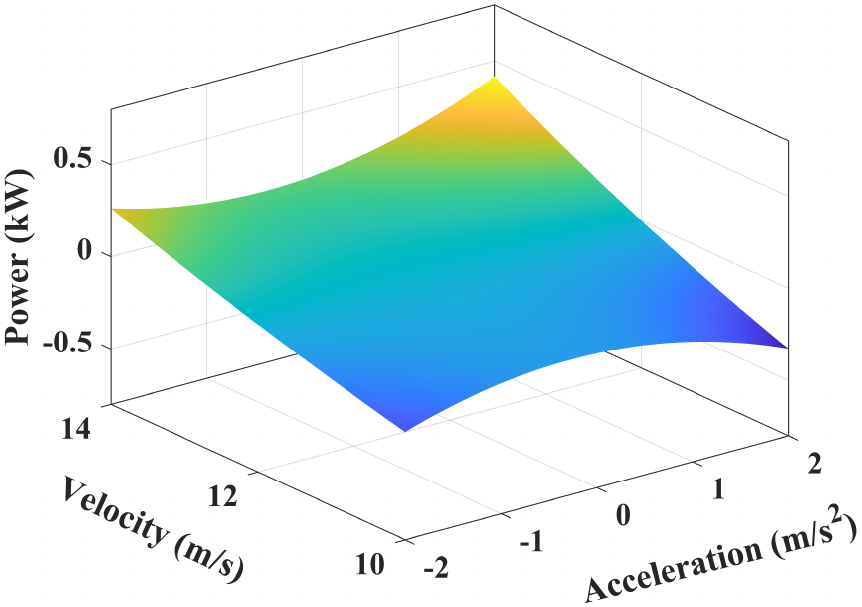}
     }
     \hfill
     \subfigure[Relative error $\frac{P_{tot}-\widetilde{P}_{tol}}{P_{tot}}$]{
         \includegraphics[width=0.225\textwidth]{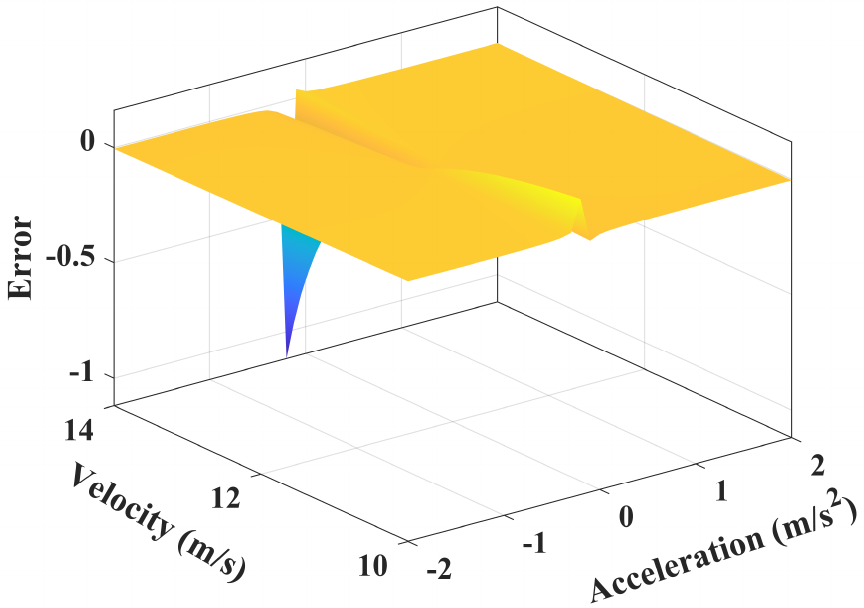}
     }
     \caption{Absolute and relative error of the estimated power under different acceleration (i.e., time-varying speed) while $\bar{v}=12 m/s$.}
     \label{fig_absolute_error}
\end{figure}

Also, the accuracy loss caused by this simplification is quantified in Fig. \ref{fig_absolute_error}, where the maximum power error is $0.42$ kW, and the relative errors are almost all $0$\% except when real power $P_{tot}$ is near $0$ kW, indicating that the estimation error can be omitted in general scenarios. 

\subsection{Characterization of Human Driver Behaviors}

In order to represent human driver behavior under the car-following scenario, the intelligent driver model (IDM) is adopted as follows:
\begin{equation}
	\label{eq_IDM}
		\dot{v}=a(1-(\frac{v}{v_d})^\delta-(\frac{s_{gap} + vT_{gap}+\frac{v\Delta v}{2\sqrt{ab}}}{s})^2), 
\end{equation}
where $\Delta v$ is the velocity difference to its preceding vehicle; $s$ is the spacing (bumper to bumper) to the preceding vehicle; definitions of other parameters are given in Table \ref{tab:IDMparameters}.

\begin{table}[htbp!]
	\caption{IDM parameters}
	\begin{center}
		\begin{tabular}{lc}
		\hline
		Parameter                        &   Value                  \\       \hline
		Maximum acceleration $a $         &   $4 m/s^2 $               \\
		Acceleration exponent $\delta$   &   $4$                    \\
		Minimum gap $s_{gap} $              &   $2 m$                  \\
		Maximum deceleration $b $         &   $5 m/s^2$                \\
		Desired velocity $v_d $           &   $25 m/s$                \\
		Time headway $T_{gap} $                 &  $(0-5) s$           \\		\hline
		\end{tabular}
	\end{center}
	\label{tab:IDMparameters}
\end{table}

\begin{figure}[!ht]
        \centering
	\includegraphics[width=3.4in]{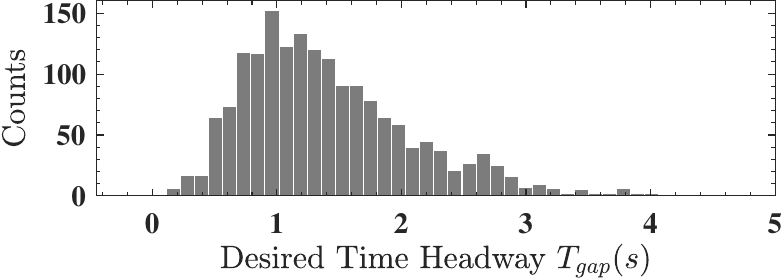}
	\caption{Distribution of desired time headway $T_{gap}$.} 
	\label{fig_Distribution}
\end{figure}

To capture the variety of human drivers, the field-test vehicle trajectory data from the NGSIM project is utilized \cite{alexiadis2004next}. By adopting the methods outlined in \cite{wang2023adaptive}, IDM is calibrated based on the car-following information of different human drivers in this NGSIM dataset. To balance optimization performance and computational efficiency, five parameters ($a_{max},\delta,s_{gap},b, v_{d}$) of IDM are set as constants, as listed in Table \ref{tab:IDMparameters}. Time headway $T_{gap}$ is selected as the main parameter to represent driver diversity, as shown in Fig. \ref{fig_Distribution}. 

\section{PA-DeePC for eco-driving of mixed traffic}\label{Application}
This section introduces the standard non-parametric system representation that characterizes mixed traffic dynamics. Subsequently, we combine PA-DeePC with Hankel matrices update, specifically designed to tackle the challenges posed by diverse, uncertain human drivers. Finally, we formulate the PA-DeePC with a novel eco-driving framework bypassing reference trajectories, and then outline its rationality and justification.

\subsection{Non-parametric Representation of mixed traffic system}

As introduced in Section \ref{Preliminaries}, the mixed traffic system will be characterized by using collected input-output trajectories. More specifically, we collect the control input sequence of CAVs, i.e., $u_{[1,T]}^{d}$, where $u^d=[u^d_{1},u^d_{2},\ldots,u^d_{m}]$, and output sequence of the following vehicles $y_{[1,T]}^{d}$, whose element is $y^{d}= [y^d_v,y^d_s]=[\widetilde{v}^d_{1},\widetilde{v}^d_{2},\ldots ,\widetilde{v}^d_{n},\widetilde{s}^d_{1},\widetilde{s}^d_{2},\ldots ,\widetilde{s}^d_{n}]$ consists of error states of velocities and spacing of CAVs and HDVs.  With those collected trajectories, the input and output Hankel matrices for non-parametric representation of mixed traffic dynamics can be constructed, as shown in Eqs. (\ref{eq_Hankel_U}), with partitioned blocks corresponding to "past" data of length $T_{ini}$ and "future" data of length $N$.

During the online implementation, at each time step $k$, a consecutive input and output trajectory of the past $T_{ini}$ steps is buffered and used to form $u_{ini}=u_{[k-T_{ini},k-1]}$ and $y_{ini}=y_{[k-T_{ini},k-1]}$. Then we define $u=u_{[k,k+N-1]}$ and $y=y_{[k,k+N-1]}$ to represent the future input and output trajectories of length $N$, which can be calculated according to the regularized non-parametric model in Eqs. (\ref{eq_regularizedcost}) and (\ref{eq_regularizedlemma}).

\subsection{Techniques for Handling Diverse HDVs}
Considering the Hankel matrices may differ a lot when facing various driving behaviors, and this may lead the DeePC fail to operate during the update process. Hence, we combined the Hankel matrix update and the proposed PA-DeePC method to improve the accuracy and adaptability. 

Initially, a consecutive trajectory with the inclusion of different and representative driving behaviors is collected to construct Hankel matrices for the unknown mixed traffic system behaviors, mainly from HDVs. This can be viewed as a more generalized Hankel matrices containing information on different driving styles. At the online adaptation stage, the PA-DeePC with this generalized Hankel matrices can handle more unknown human driving behaviors since such Hankel matrices have some similar and short trajectories of the real-time HDVs. Then, to enhance the adaptability to specific and unknown human drivers, we update the Hankel matrices online by including trajectories of the current trajectory.  

 For mixed traffic system, we can find the following physics: (1) The relationship between velocity and acceleration (${v}_{i,k+1} - {v}_{i,k} = a_{i,k} \Delta t$) of one CAV; (2) The relationship between the spacing of following vehicles and their velocities and accelerations (${s}_{i,k+1} - {s}_{i,k} = ({v}_{i-1,k} - {v}_{i,k})\Delta t + \frac{1}{2}(a_{i-1,k}-a_{i,k})(\Delta t)^{2} $).
 When the system is near equilibrium state ($a_{i-1,k}-a_{i,k} \approx 0$) or the sampling time step $\Delta t$ is small enough,  $\frac{1}{2}(a_{i-1,k}-a_{i,k})(\Delta t)^{2}\approx 0$. Hence, this physics can be rewritten as follows to construct the proposed PA-DeePC
\begin{equation}
	\label{eq_PAconstraints}
		\begin{aligned}
			& \epsilon_{1,k} = \widetilde{s}_{i,k+1} - \widetilde{s}_{i,k} - (\widetilde{v}_{i-1,k} - \widetilde{v}_{i,k}) \Delta t ,& i \in \Omega \setminus S_{1} ,     \\ 
			& \epsilon_{2,k} = \widetilde{v}_{i,k+1} - \widetilde{v}_{i,k} - a_{i,k} \Delta t, & i \in S,
		\end{aligned}
\end{equation}
where $S_{1}$ denotes the first CAV in set $S$. 

Assuming that we have already established the cost function and the corresponding constraints for mixed traffic, the following procedures is used to learn and predict diverse human driver behaviors.

\begin{algorithm}
	\begin{algorithmic}[1]
	\State \textbf{Initialize} generalized $U_{p}, U_{f}, Y_{p}, Y_{f}$ \label{algorithm}
	\Procedure{Online adaptation}{New Behavior}
	\For{$i \gets 1, n$}
		\State \textbf{run} \textit{PA-DeePC} and collect ($u_{k},y_{k}$)
		\State Get $U_{p}^{'}, U_{f}^{'}, Y_{p}^{'}, Y_{p}^{'}$ with ($u_{k},y_{k}$)
		\If{$rank(U^{'}) == rank(U) $}
			\State $U_{p}, U_{f}, Y_{p}, Y_{f} \gets U_{p}^{'}, U_{f}^{'}, Y_{p}^{'}, Y_{f}^{'}$
		\EndIf
		\State \textbf{end}
	\EndFor
	\EndProcedure
	\Procedure{Online implementation}{Test Cycle}
	\State \textbf{Initialize} particular $U_{p}, U_{f}, Y_{p}, Y_{f}$
        \For{$i \gets 1, n$}
	   \State \textbf{run} \textit{ PA-DeePC}
        \EndFor
        \State \textbf{end}
	\EndProcedure
	\end{algorithmic}
	\caption{Adaptive PA-DeePC algorithm}
\end{algorithm}

\subsection{Formulation of PA-DeePC for Eco-driving}

In this study, the objective is to control CAVs to operate within a suitable spacing that is neither too small, leading to collisions, nor too large, reducing traffic efficiency, while minimizing energy consumption when HDVs have diverse driving behaviors. 

The conventional logic to achieve this objective is, using a pre-defined reference $(v^{*},s^{*})$ to regulate the trajectory, adding a power term into the cost function to optimize energy consumption, and then tuning weighting parameters to balance the trade-off between the above two cost terms. However, pre-defined speed and spacing reference trajectories may be helpful when system dynamics are deterministic \cite{yamashita2016reference}. Yet, they can limit optimization performance in situations where system dynamics vary due to diverse human drivers. Particularly, it is unlikely that following specific velocity and spacing references can achieve optimal eco-driving performance, especially in dynamic traffic environments. Furthermore, controlling a CAV at a significant distance from the PV and following the velocity $(v^{*})$ of the PV is deemed unreasonable, particularly when sudden and significant changes occur. In consequence, we propose a novel eco-driving framework of mixed traffic that bypasses specific trajectory references to give larger optimization space, as shown below.
\begin{equation}
	\label{eq_cost2}
        \begin{aligned}
            \underset{g,u,y,\sigma_y}{min} & \sum_{k=0}^{N-1}(\Vert \widetilde{P}_{k} \Vert_S + \Vert u_k \Vert^2_R + \Vert \epsilon_{k} \Vert^{2}_{M})  +  \lambda_g\Vert g \Vert^2_2 + \lambda_y\Vert \sigma_y \Vert^2_2 , \\
            \text{s.t.} \quad & (\ref{eq_regularizedlemma}) ,   \\
            & u_{k} \in [a_{min},a_{max}], \quad \forall k \in \{0,\cdots,N-1\}, \\
		& y_{s,k} \in [\widetilde{s}_{min},\widetilde{s}_{max}], \quad \forall k \in \{0,\cdots,N-2\}, \\ 
		& y_{s,k} \in [\widetilde{s}_{safe},\widetilde{s}_{effi}], \quad k=N-1,\quad i \in S, 
        \end{aligned}
\end{equation}
where $\widetilde{P}_{k}$ represents the estimated total power; $M$ is the penalty matrix for residual of considered physics; $a_{min/max}$ represents the lower and upper bounds of control inputs; while $\widetilde{s}_{min/max}$ and $\widetilde{s}_{safe/effi}$ are loose and tight bounds for CAVs, respectively, which can be represented by the constant time headway (CTH) $t_{h}$, a safety metric \cite{chen2019effects}, and the time gap ($TG$), a traffic efficiency metric \cite{wang2023adaptive}, as follows
\begin{equation}
	\label{eq_CTHTG}
	\left\{
		\begin{aligned}
			& \widetilde{s}_{min/safe} + s^{*} = v_{f}t_{h} + s_{gap}, \\
			& \widetilde{s}_{max/effi} + s^{*}= v_{f}TG,
		\end{aligned}
	\right.
\end{equation} 
where the velocity of the preceding vehicle $v_{i-1}$ is utilized as $v_f$ for vehicle $i$. Compared with Eq. (\ref{eq_regularizedcost}), without the balance of the trajectory term $(\Vert y_{k} \Vert _{Q}^{2})$, the power term $(\Vert P_{k} \Vert _{S}^{2})$ could cause CAVs slow down as much as possible for reduced energy consumption in the short term; however, traffic efficiency can be dramatically impaired. This is caused by the limited prediction horizon in almost all predictive control algorithms, thus the long-term benefit and optimality cannot be fully considered.  
To address this issue, we combine loose and tight constraints on spacing, where the loose constraints (a smaller $t_{h}$, and a larger $TG$) for the short term give larger optimization space, and the tight ones (a larger $t_{h}$, and a smaller $TG$) for the long term avoid the excessive slowdown effect of the power term by providing a smaller spacing range.

\section{Simulation results}\label{Simulation}
This section presents the implementation and evaluation of the proposed PA-DeePC approach for eco-driving of mixed traffic flows considering diverse human behaviors.  

\subsection{Experimental Setup}

In this study, a 5-vehicle platoon ($S=\{1,3\}$) system model is created in MATLAB to accurately capture traffic information. Based on this model, a generalized consecutive trajectory ($T=1000$) is collected with the sampling interval of $\Delta t = 0.1$s. To incorporate diverse human driver behaviors, the time headways of two HDVs are gradually and randomly changed based on the distribution shown in Fig. \ref{fig_Distribution} every $50$ steps. Subsequently, we construct generalized Hankel matrices using the data collected offline.

The parameters used in this study are described as follows. The time horizons for the future and past trajectories are set to $N = 40$, and $T_{ini} = 20$, respectively. Constraints are imposed on acceleration and its values specified in Table \ref{tab:IDMparameters}. Then, $t_{h} = 1/1.3$s and $TG= 3.5/2.1$s are used to calculate $s_{min/safe}$ and $s_{max/effi}$, respectively. At last, weighting parameters in the cost function (\ref{eq_cost2}) are given as follows $S = 0.1$, $R = 1$, $M = 500$, $\lambda_{g} = 20$, and $\lambda _{y}=1000$ \cite{coulson2019data}. 

The test speed profile for the PV spans six minutes with a sampling rate of $0.1$s, using the NGSIM data set as to represent real-world driving scenarios. A stochastic test including random HDVs is selected to verify the adaptability of our approach in predicting different human driver behaviors. To ensure a thorough and representative test of human driving diversity, we select a range of time headways from the time headway distribution shown in Fig. \ref{fig_Distribution}. The primary focus is on time headways in the range of $T_{gap} \in [0.5,2.9]$s \cite{wang2023adaptive}, divided into 12 equally spaced groups with a $0.2$s interval.
Then, we randomly select two distinct human drivers (i.e., two random IDM headway parameters) from each group to create a test set that accurately reflects the majority of human driving behaviors. To represent almost all possible human driving behaviors, we consider all combinations of the test set, resulting in 576 scenarios. Then, we apply the overall simulation stage pipeline outlined in Algorithm \ref{algorithm} to each combination.

To provide a fair comparison, we establish two baselines, one implementing the PA-DeePC with a conventional optimization framework, where the weighting parameters for reference ($\Vert y_{k} - r\Vert^2_{Q}$) are the same as those in \cite{wang2023deep} ($Q_{v} = 1, Q_{s} =0.5$); another one implementing the OVM-based adaptive cruising control (ACC) algorithm with two identical parameter sets. Specifically, we set $\alpha = 0.8,\beta = 0.5$ as an aggressive human driver, and $\alpha = 0.5,\beta = 0.8$ as a mild human driver \cite{zheng2020smoothing}. The simulations are executed in MATLAB R2022b on a Windows 11 PC with a 3.8GHz processor and 32GB of RAM.

\subsection{Performance of PA-DeePC}
The simulation study aims to evaluate the effectiveness and adaptability of the proposed PA-DeePC and the eco-driving optimization framework. The system characterization and prediction performance comparison between the original DeePC and the PA-DeePC is firstly given to demonstrate the improvement of adding physics. Then, the energy savings performance of the proposed optimization framework is assessed based on the benchmark results. At last, safety and traffic efficiency in vehicular traffic are briefly introduced.

\begin{figure}[!htpb]
	\centering
	\subfigure[Prediction spacing and velocity error during online update]{
		\includegraphics[width=3.4in]{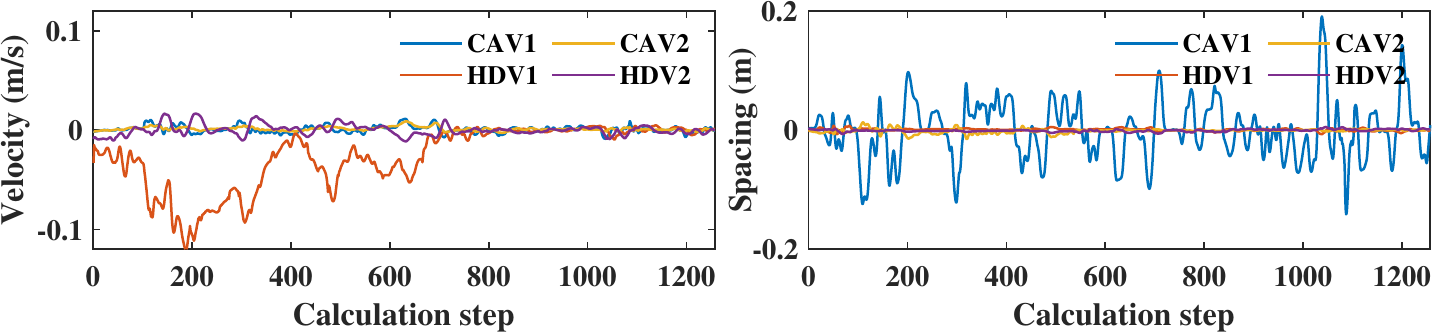}
	}
	\hfill
	\subfigure[Prediction error comparison among the original DeePC and PA-DeePC]{
		\includegraphics[width=3.4in]{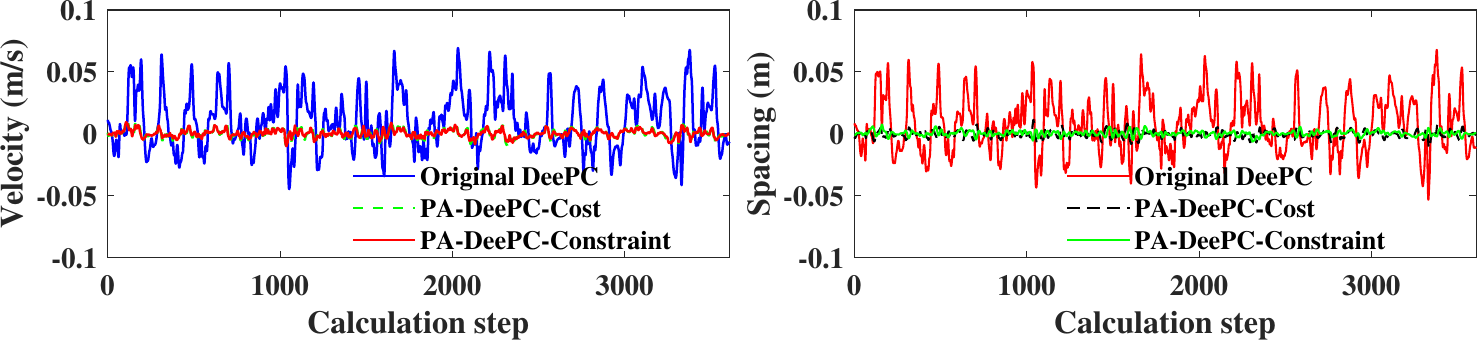}
	}
	\caption{Prediction error during of update and online implementation. The errors are computed via $e_{v} = \widetilde{y}_{v} - \widetilde{v}$, and $e_{s} = \widetilde{y}_{s} - \widetilde{s}$, respectively.}
	\label{fig_PredError}
\end{figure}

\textbf{System Characterization and Prediction:}
To evaluate the performance of incorporating physics and updating Hankel matrices, we can use the predicted velocities and spacing of CAVs and following vehicles except CAV1 as metrics. Fig. \ref{fig_PredError} shows the prediction errors for the $k+1$ step. Particularly, Fig. \ref{fig_PredError}a depicts the prediction errors of following vehicles of PA-DeePC can converge to zero with real-time updates of Hankel matrices, while the velocity errors of CAVs can always maintain in small range due to those incorporated physics. Further, to directly demonstrate the difference of adding physics into the original DeePC, we depict the comparison results in one case in Fig. \ref{fig_PredError}b. In this comparison, both the original DeePC and the PA-DeePC operate using the updated Hankel matrices. According to the simulation results, even though the original DeePC demonstrates the ability to effectively capture system dynamics with relatively high accuracy $(\leq 0.06)$, the PA-DeePC shows superior performance in predicting the system dynamics $(\leq 0.01)$. With that being said, adding physics can significantly improve the ability of DeePC in capturing system dynamics.

\begin{figure}[!ht]
	\centering
	\includegraphics[width=3.4in]{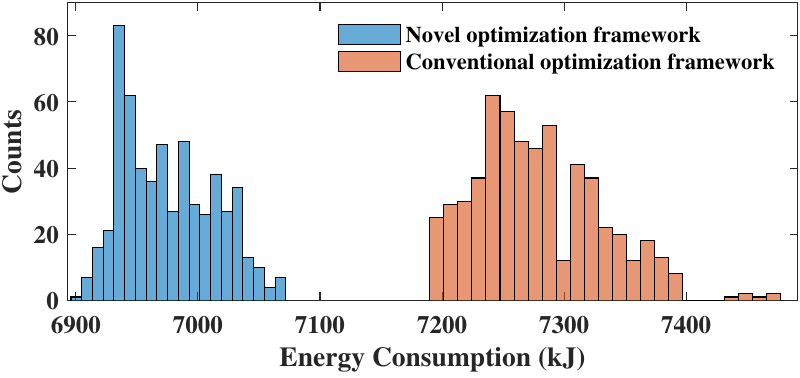}
	\caption{Comparison of the energy consumption distribution between two different optimization frameworks under the influence of measurement noise.} 
	\label{fig_EnergyConsumption}
\end{figure}

\textbf{Eco-driving Performance:} The eco-driving performance is evaluated by examining the energy consumption of all vehicles across 576 simulation cases. Furthermore, when accounting for potential measurement inaccuracies in real-world applications, we introduce a Gaussian noise term $(w \in [-0.1, 0.1])$ into both the real-time spacing and velocities. Next, we present a comparison between the novel optimization framework and the conventional optimization framework, as depicted in Figure \ref{fig_EnergyConsumption}. The results demonstrate a notable reduction in energy consumption $(4.07\%)$ when compared to this baseline, underscoring the superior optimization achieved by the novel framework. Additionally, the energy consumption remains relatively stable, indicating that the proposed approach can effectively handle diverse human driver characteristics and deliver stable results. Then, Table \ref{tab:EnergyConsumption1} provides a detailed comparison for each vehicle compared with two baselines. Specifically, when compared to the aggressive OVM, the average energy reduction achieved by the proposed approach is $9.29\%$. For the mild OVM, the average energy consumption reduction attained by PA-DeePC is $4.97\%$. These findings highlight the potential of the proposed approach to enhance energy efficiency in mixed traffic.

\begin{table*}[!htpb]
	\centering
	\caption{Platoon energy consumption under diverse driving characteristics}
	\begin{tblr}{
	  cell{1}{2} = {c=7}{c},
	  hline{1,3,8} = {-}{},
	  hline{2} = {2-8}{},
	}
		  & Energy consumption (Average) (kJ) &               &                   &               &                  &               &                  \\
		  & PA-DeePC                & Original DeePC & Improvement & OVM($\alpha = 0.8,\beta = 0.5$) & Improvement & OVM($\alpha = 0.5,\beta = 0.8$) & Improvement \\
	CAV1  & 1788.97          & 1998.23     & 10.47\%           & 1919.95     & 6.82\%           & 1886.48     & 5.17\%           \\
	HDV1  & 1751.59          & 1807.26     & 3.08\%            & 1895.27     & 7.58\%           & 1824.67     & 4.01\%           \\
	CAV2  & 1728.48          & 1742.90     & 0.83\%            & 1918.97     & 9.93\%           & 1814.87     & 4.82\%           \\
	HDV2  & 1706.86          & 1728.71     & 1.26\%            & 1956.10     & 12.74\%          & 1814.89     & 5.95\%           \\
	TOTAL & 6975.90          & 7277.09     & 4.14\%            & 7690.29     & 9.29\%           & 7340.91     & 4.97\%           
	\end{tblr}
	\label{tab:EnergyConsumption1}
\end{table*}

\begin{figure}[!htpb]
	\centering
	\includegraphics[width=3.4in]{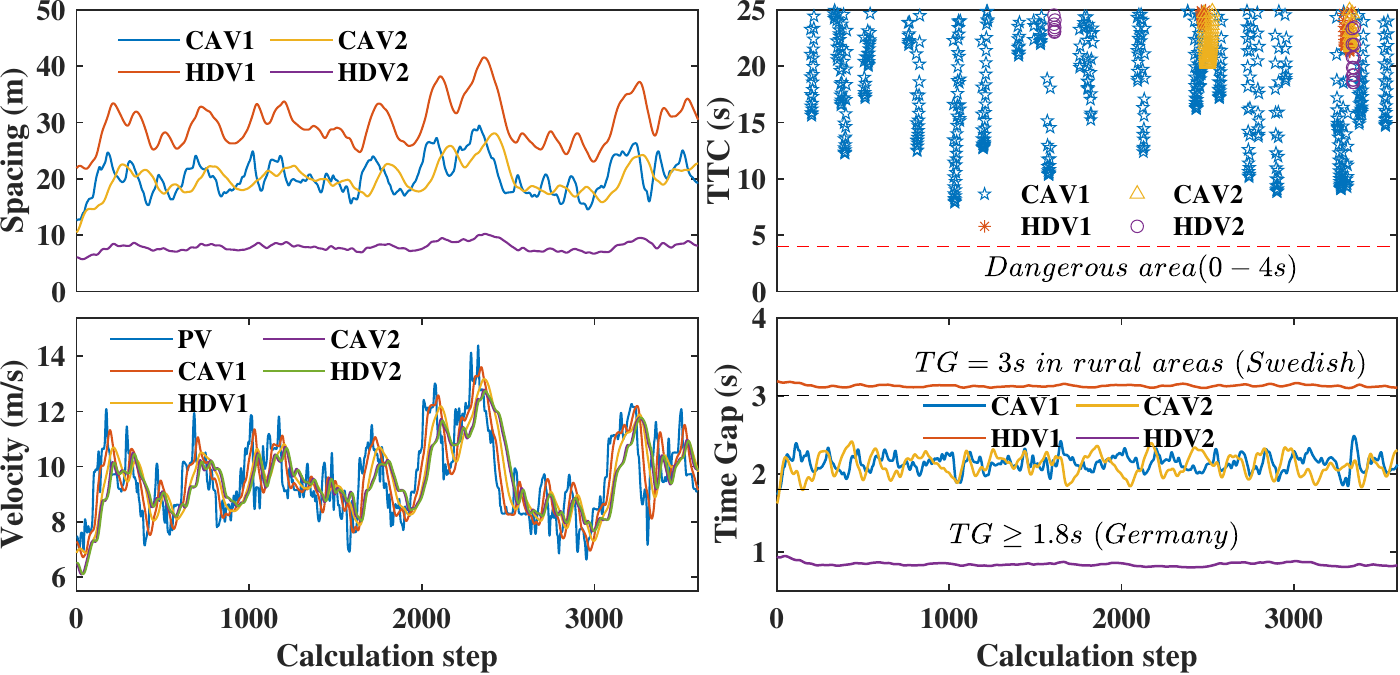}
	\caption{Traffic information of $s$, $v$, $TTC=-\frac{s_{i}}{\Delta v_{i-1,i}}$, $TG = \frac{s_{i}}{v_{i}}$. } 
	\label{fig_TrafficInfo}
\end{figure}

\textbf{Safety and Traffic Efficiency:} With such great energy saving performance, the safety and traffic efficiency of the proposed optimization framework are investigated by analyzing traffic information, as shown in Fig. \ref{fig_TrafficInfo}. Two commonly employed metrics: time-to-collision ($TTC$) and time gap, are chosen for examination. $TTC$ metric quantifies the time left until a potential collision occurs and is considered a more intuitive metric for identifying potential danger compared to the CTH metric. In this study, we set the criteria of $0\leq TTC \leq 4$s and $TG \geq 3$s, as shown in Fig. \ref{fig_TrafficInfo}, to identify unsafe and inefficient conditions, respectively. The details of setting both metrics can be found in \cite{wang2023adaptive,vogel2003comparison,zhu2020safe}.
 The simulation results demonstrate the proposed PA-DeePC can effectively avoid dangerous driving under such aggressive driving scenarios, as evidenced by the fact that the most critical point of CAV1 still have $TTC > 7.47$s. In addition, simulation results suggest that the time gaps for both CAVs are regulated within a range of $1.5-2.6$s, ensuring optimal traffic efficiency with enough safety margin.

\section{Conclusion}\label{Conclusion}
In this paper, we introduce an innovative Physics-Augmented Data-EnablEd Predictive Control method and a novel optimization framework for eco-driving of mixed traffic flows with diverse human driver behaviors. By incorporating partial system physics into the original DeePC, the ability to characterize system dynamics is significantly improved. Specifically, the proposed framework can accurately characterize human driver behaviors, and the novel optimization framework can decently reduce the holistic energy consumption of both CAVs and HDVs by $4.14\%$ on average while ensuring driving safety and traffic efficiency. In addition, simulation results under a variety of random traffic environments show that the proposed optimization method can reduce the energy consumption of mixed traffic flows by $4.97\%$ and $9.29\%$ on average when compared to benchmark results obtained by implementing the OVM-based cruising control algorithm with two sets of parameters, respectively.

\ifCLASSOPTIONcaptionsoff
  \newpage
\fi

\bibliographystyle{IEEEtran}
\bibliography{refs}

\end{document}